\documentclass{aa}
\usepackage{graphicx}
\usepackage{txfonts}
  \def\etal{et al.}

\begin{document}
   \title{JEM-X background models\thanks{Based on observations with
    INTEGRAL, an ESA project with instruments
    and science data centre funded by ESA member states (especially the PI
    countries: Denmark, France, Germany, Italy, Switzerland, Spain), Czech
    Republic and Poland, and with the participation of Russia and the USA.}
}

   \author{J. Huovelin, \inst{1}
           S. Maisala, \inst{1}
           J. Schultz, \inst{1}
           N. J. Westergaard, \inst{2}
           C. A. Oxborrow, \inst{2}
           P. Kretschmar, \inst{3,4}
           \and
           N. Lund \inst{2}
          }
   \offprints{J. Huovelin, \\ email: Juhani.Huovelin@Helsinki.Fi}

   \institute{Observatory, P.O.B. 14 (Kopernikuksentie 1),
	      FIN-00014 University of Helsinki, Finland
              \and
              Danish Space Reasearch Institute,
              Juliane Maries Vej 30,
              DK 2100 Copenhagen, Denmark
              \and
              Max-Planck-Institut f\"ur Extraterrestrische Physik, 
              Giessenbachstrasse, 85748 Garching, Germany
	      \and
              INTEGRAL Science Data Center, Chemin d'Ecogia 16, 
              Versoix, Switzerland
             }

   \date{Received July 15, 2003; accepted }

\authorrunning{Huovelin \etal}

\abstract{
Background and determination of its components for the JEM-X 
X-ray telescope on INTEGRAL are discussed. A part of the first background
observations by JEM-X are analysed and results are compared
to predictions. The observations are based on extensive
imaging of background near the Crab Nebula on revolution 41 of 
INTEGRAL. Total observing time used for the analysis 
was 216502 s, with the average of 25 cps of background for 
each of the two JEM-X telescopes. JEM-X1 showed slightly higher
average background intensity than JEM-X2. The detectors were
stable during the long exposures, and weak
orbital phase dependence in the background outside radiation belts was 
observed. The analysis yielded an average of 5 cps for the
diffuse background, and
20 cps for the instrument background. The instrument background was found
highly dependent on position, both for spectral shape and
intensity. Diffuse background was enhanced in the central area
of a detector, and it decreased radially towards the edge, with a
clear vignetting effect for both JEM-X units.
The instrument background was
weakest in the central area of a detector and showed a steep
increase at the very edges of both JEM-X detectors, with significant
difference in spatial signatures between JEM-X units. According
to our modelling, instrument background dominates over diffuse background
in all positions and for all energies of JEM-X. 

      \keywords{X-ray background --
                X-ray data analysis --
                INTEGRAL satellite
               }
} 

%

   \maketitle

\section{Introduction}

   \begin{figure}
 \resizebox{\hsize}{!}{\includegraphics{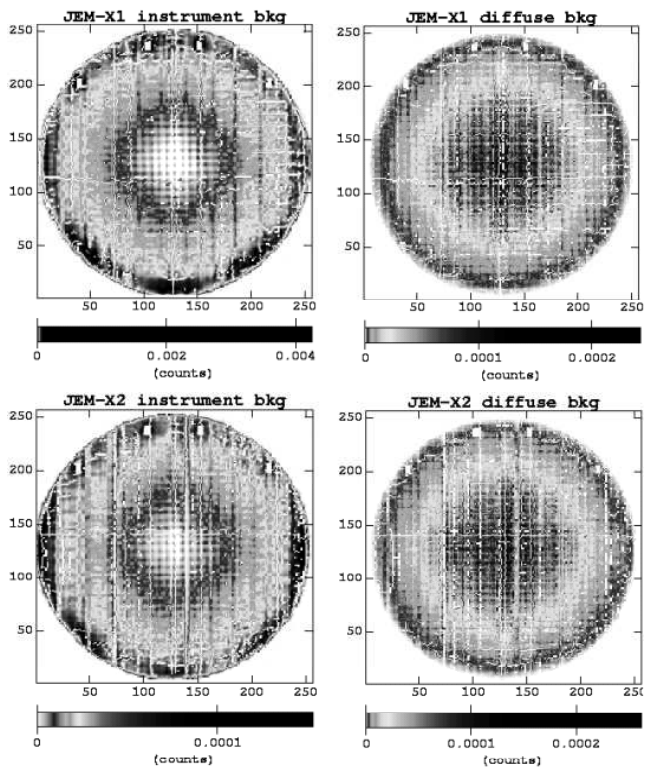}}
     \caption{Spatial distribution of the background. Upper panels:
     JEM-X1, lower panels, JEM-X2. Left panels: Instrument
     background, Right panels: Diffuse background. The white
     rectangles denote the positions of the calibration
     sources, which have been excluded from our analysis. 
     The collimator signature can be seen as weak vertical and
     horizontal line structures in the shadowgrams. The broad
     vertical lines are due to dead anodes. Also some photon
     leak from the calibration sources is evident. The sharp and very 
     narrow lines are graphical artifacts caused by the plotting
     routine. Total intensity of each shadowgram is normalized to 1.
         \label{Fig_bkg_distr}}
    \end{figure}
%

Background of X-ray and $\gamma$~-ray detectors for astronomy 
can generally be divided in two components, {\em diffuse sky background} and 
{\em instrument background}. 

JEM-X consists of two mechanically identical telescope units (JEM-X 1
and 2), with a position sensitive xenon-filled Microstrip Gas Chamber,
a collimator, and a coded mask as an optical element in each unit  
(see Lund \etal~ 2003 for more details). There are four internal 
radioactive sources for spectral calibration in the collimator of each unit. 

The sky image is thus a result of mathematical processing of the mask
shadow pattern on the position sensitive detectors (e.g. in't Zand 1992;
in't Zand, Heise \& Jager 1994).

{\em Diffuse sky background} enters the detectors via the aperture and 
is affected by the mask and the collimator,
similarly to the distinguishable sources in the sky. 
{\em Instrument background} includes detector signal
due to interactions between cosmic radiation and materials of the satellite 
and the detector itself.  MeV-range or higher energy $\gamma$-ray
detectors  may also be sensitive to direct $\gamma$-rays from e.g. 
solar flares that penetrate the detector through the spacecraft
and detector panels (e.g. Ferguson \etal, 2003). Instrument
background thus carries essentially no coding from the mask. It 
introduces additional noise and may bring up artificial
sources, if not properly determined in the shadowgrams, and then
accounted for in image reconstruction. 
Useful examples of early treatment and 
analysis of instrument background for coded aperture telescopes
can be found from Covault \etal~(1991) and Willmore \etal~(1992).
 
Due to the expected difficulty of background determination, the JEM-X 
instrument specific software (ISSW) has been designed to 
permit accounting for various effects and dependencies of
background, and background modelling is made using external
background libraries (see Westergaard \etal, 2003). 
Detailed description of the software can be found in the 
Architectural Design Document (ADD) of JEM-X (Oxborrow \etal, 2002).

In this paper, we present results and analysis of
a part of the first extensive
background observations with INTEGRAL JEM-X.

%
   \begin{table}
      \caption[]{INTEGRAL background pointings during cycle 41.}
\begin{center}
\begin{tabular}{ccc}
            \hline
    $\alpha$(2000.0) & $\delta$(2000.0)  & staring time  \\
      (h~m~s) & ($^{\circ}$~$'$~$''$)  & (s) \\
            \hline
            4 53 24.0 & $+$21 08 50.1 & 70000  \\
            5 36 45.0 & $+$12 24 15.1 & 40825  \\
            5 36 45.2 & $+$12 25 46.4 & 42775  \\
            5 36 45.3 & $+$12 25 43.5 & 64468  \\
            \hline
\label{tab1}
\end{tabular}
\end{center}
\end{table}

\section{Observations}

Observations included in our analysis were made during cycle 41 of 
INTEGRAL, between February 13, 01:05:57 UTC, and February 16, 00:51:18
UTC, 2003. The pointings for these background observations were 
selected from the vicinity of the Crab Nebula, with pointing centres about 
$9^{\circ}$ away from the position of Taurus XR-1 (see Table 1).
The total observing time of JEM-X was 218068 s, of which 216502.12 s 
was included in our analysis. The background observing programme 
near Crab Nebula started already during cycle 40. The first 
observations were not included
in our analysis, since we aimed at studying the background and the
variations within one full orbit. JEM-X observations were made in full
imaging mode. We used JEM-X events preprocessed with standard
JEM-X ISSW, which is publically distributed by the ISDC. 
The spatial gain corrections and other preprocessing
parameters date to May 20, 2003.

\section{Analysis}

The analysis included fitting of a two component background model to 
JEM-X spectra for 8 radial ranges of equal area beginning from
the centre of each of the two detectors. The radial ranges
were further divided in 6 azimuthal sectors each covering an angle of
$60^{\circ}$ (Fig. 3). 
The modelling was made using the publically available XSPEC
X-ray spectrum fitting software (Arnaud 1996). The energy range used
for the spectral fits was 4-33 keV. 

The following assumptions were made
for the fitting procedure: First, we assumed the following
analytical formula for the diffuse sky background,
\begin{equation}
I = C E^{-1} \mathrm{e}^{-E/ 40 \, \mathrm{keV}} 
\mathrm{photons} \,  \, \mathrm{cm}^{-2} \,  
\mathrm{s}^{-1} \, \mathrm{keV}^{-1}
\end{equation} 
where $E$ is photon energy and C is a normalization constant.
This is close to a $\sim 40 \mathrm{kev}$ thermal bremsstrahlung model 
well fitted for diffuse X-ray background from 3 to 45 keV (e.g.
Marshall \etal 1980). 
Second assumption was that, excluding the known line emission
from the detector and surrounding material,
instrument background should be flat (i.e. constant in energy).
The lines were modeled as narrow (maximum width determined by instrumental
energy resolution) Gaussian lines with fixed line centroid positions.
Eleven lines were detected in the background spectra.
Thus, the background spectral model
has 24 free parameters, including widths and strengths of the eleven
lines and normalization of the two continuum components. 

The modelling was made in two steps. First, we fitted the spectra
extracted from each subregion of both detectors
with the full model described above to get an initial estimate for the
diffuse background level. After this, the estimated diffuse
background component was subtracted from each spectrum,
and the residual spectrum was taken as the instrument background.

The resulting background images and sample spectra of a few
subregions are shown in Figs. 1 and 2.
For a summary of the results, see Tables 2 and 3.
All parameters for the background components
in Tables 2 and 3 are directly from the XSPEC spectral fitting, which
applies instrument response and effective area similarly to
all model components. 
The fitting procedure thus yields physically consistent values only for the
diffuse background component, which is modified by the coded mask and 
collimator. As for the instrument background, however, full mask coding
and effect of the collimator are not relevant assumptions.
Thus, the flux values for the instrument background in Table 2 are not
physically valid, but still usable for the modelling purpose.
This should be kept in mind when using the values from Tables 2 and 3.

The diffuse background decreases towards the edges
of the detector, as expected.
The instrument background is stronger than expected, dominating 
the spectrum at all radii.
The ten expected K-shell lines (from the 
$^{109}$Cd and $^{55}$Fe calibration sources, collimator (Mo),
and detector gas (Xe)) were detected close to their
nominal positions. This implies that the energy scale is correctly
determined. The previously unknown weak line near 13 keV turned out to be
the uranium L-shell line. It most likely
originates in the detector beryllium window.
Near the edges of the detector, the background is
highly nonuniform. Additional nonuniformity in the outer parts was
introduced by photon leak from calibration sources, which could not
be completely eliminated.

\begin{table}
\caption{The normalization factors of the background continuum components
 ($10^{-3} \mathrm{photons} \, \mathrm{keV}^{-1} \, \mathrm{cm}^{-2} \,  
  \mathrm{s}^{-1}$ at 1 keV). Mean and standard deviation of values
 derived from the six extraction regions at each radius are given.
  The energy range used in the fitting is 4-33 keV. 
  {\em Diffuse} denotes diffuse sky background, {\em Flat}
  denotes the flat continuum of the instrument background. Note that the
normalization is determined on the basis of
source spectra from $\frac{1}{48}$ of each detector area.}
\begin{tabular}{cccccc}
\hline
$R_{\mathrm{in}}$ & $R_{\mathrm{out}}$ & \multicolumn{2}{c}{JEM-X1} & 
\multicolumn{2}{c}{JEM-X2} \\
pix      & pix       & Diffuse & Flat  & Diffuse & Flat \\
\hline
0    & 47.6  & $75 \pm   6$   &  $4.34 \pm 0.11$ 
& $68 \pm    13$ & $4.06 \pm   0.27$ \\
47.6 & 64.8  & $62 \pm   5$ &  $3.83 \pm 0.24$
& $64 \pm   12$ & $3.73 \pm   0.31$ \\
64.8 & 78.4  & $60 \pm  8$ &  $4.15 \pm 0.22$ 
& $64 \pm 11 $ & $4.00 \pm   0.32$ \\
78.4 & 90    & $53 \pm   4$ &  $4.39 \pm 0.21$
& $61 \pm    11$ & $4.16 \pm   0.31$ \\
90   & 100   & $47 \pm   6$ &  $4.88 \pm 0.24$ 
& $55 \pm   7$ & $4.80 \pm   0.60$ \\
100  & 109   & $24 \pm    15$ &  $6.57 \pm 1.11$  
& $51 \pm    12$ & $5.66 \pm   0.45$ \\
109  & 118   & $0 \pm      0$ &  $9.4 \pm  2.3$ 
& $30 \pm   9$ & $8.3 \pm    2.2$ \\
118  & 126   & $9 \pm    14$ &  $7.7 \pm  1.5$ & 
$17 \pm 13$ & $7.7 \pm  3.0$ \\
\hline
\end{tabular}
\end{table}

   \begin{figure}
\resizebox{\hsize}{!}{\includegraphics{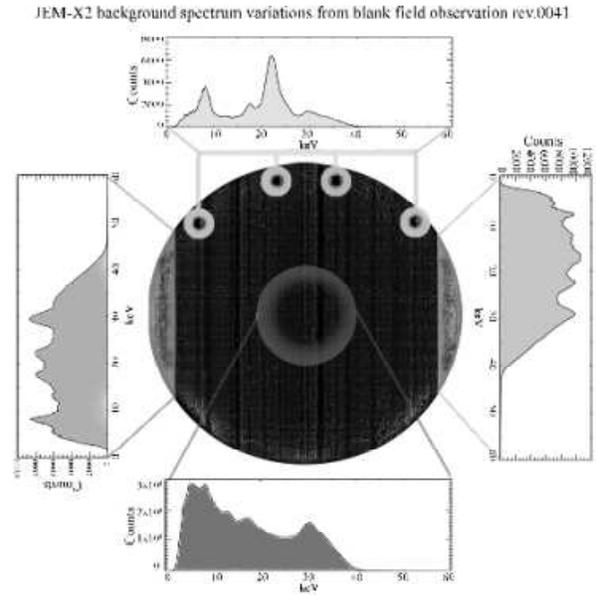}}
     \caption{Four sample background spectra extracted from
     different parts of JEM-X2. At the sides of the detector,
     a blend of K-shell lines from the spacecraft structure is seen. 
     Note also the prominent lines in spectrum extracted from the
     surroundings of the calibration sources.        
 \label{Fig_bkg_sp}}
    \end{figure}

   \begin{figure}
\resizebox{\hsize}{!}{\includegraphics{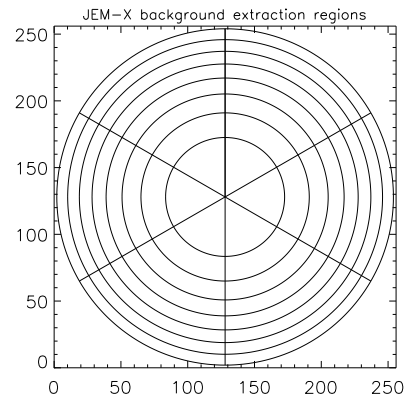}}
     \caption{The background extraction regions. Units in both axes
     are pixels. All regions cover an
     equal area of the detector. Exclusion of calibration sources
     (not shown) reduces the actual area of some regions.
      \label{Fig_piechart}}
    \end{figure}

\begin{table}
\caption{The lines detected from the background. Line ID is the
element and transitions producing the line,  $E$ is the 
line energy in keV (Thompson et al., 2001). 
Subscripts 1 and 2 denote the detectors JEM-X1 and 2, $F$ is the
largest line strength detected, $\bar{F}(N)$ mean of detected line
strengths where $N$ is the number of regions from which
the line is detected (maximum is 48 regions / detector).
The Mn/Fe line at 6.45 keV is a blend of  Mn K$_\beta$ (6.49 keV)
and Fe K$_\alpha$ (6.40 keV). Line strengths are given 
in $10^{-3} \mathrm{photons} \, \mathrm{cm}^{-2} \, \mathrm{s}^{-1}$.
Note that the line strengths are determined on the basis of
source spectra from $\frac{1}{48}$ of each detector area,
and the tabulated line strengths are nominal results of
XSPEC fitting. As the fitting is done in energies between 4-33 keV, the flux
values of Xe K$_\beta$ line at 33.6 keV are based on extrapolations.}
\begin{tabular}{lcccccl}
\hline
Line&$E$&$F_1$&$\bar{F_1} (N)$&$F_2$&$\bar{F_2}(N)$&Origin\\ 
\hline
Mn K$\alpha$ & 5.90 & 9.0 &  8.1 (2) 
& 4.5 & 2.5 (34)& $^{55}$Fe \\
Mn/Fe & 6.45 & 15.4 &  15.4 (1)   
& 2.4 & 1.8 (3) & $^{55}$Fe \\
Ni K$\alpha$ & 7.47 & -       &  -          
& 3.4 &  2.2 (6) & $^{109}$Cd \\
Ni K$\beta$ & 8.27 &  12.4 & 9.6 (38)  
&  30.5 & 15.5 (48) & $^{109}$Cd  \\
U L$\alpha$& 13.5 & 22.2 & 8.6 (27) &
58.6 &  11.5 (42) & Be win \\
Mo K$\alpha$& 17.4 & 41.9 & 9.2 (43)  
& 69.8 & 14.0 (47)& Collim. \\
Ag K$\alpha$& 22.1 & 19.7 & 11.2 (7) 
& 51.1 & 15.2 (19) & $^{109}$Cd\\
Cd K$\alpha$& 23.0 & 12.3 &  10.9 (2)
& 15.0 &  3.6 (4) & $^{109}$Cd\\
Ag K$\beta$& 24.9 & 16.5 &  11.5 (5)
& 13.6 & 7.7 (13) & $^{109}$Cd \\
Xe K$\alpha$& 29.6 & 46.1 & 14.7 (40)
& 64.0 & 19.2 (48)  & Gas \\
Xe K$\beta$& 33.6 & 31.0 &  14.0 (24) 
& 25.9  &  11.6 (31)& Gas \\
\hline
\end{tabular}
\end{table}

We also searched for possible dependence of background on 
the orbital phase of the observations. The spectrum varied
with a range of approximately 5\% between three separate orbital
sections well outside radiation belts. The variation is statistically 
significant but small.
Also dependence on solar aspect angle and
particle radiation level can be utilised in the JEM-X
ISSW background modelling. Significant variations were not found.  
The variation in the solar aspect 
due to different pointings was $20^\circ$, which is probably 
not sufficiently
large for studies of an effect on instrument background. Also, there was
no proper indicator of particle radiation level on INTEGRAL
available during our observations to search for a correlation.

\section{Conclusions}

We have analyzed a part of the first INTEGRAL background observations
with JEM-X. Estimates of the spatial and spectral
distributions are obtained
for diffuse sky background and instrument background.

The total background observed for JEM-X1 was 28~cps, for JEM-X2
23~cps, and 25~cps on the average. A part 
($\sim 1/5$) of the excessively
large background may be due to residual Crab Nebula emission in JEM-X data.

According to XSPEC fitting, the diffuse background was at 
maximum in the centre of the detector and it decreased
radially towards the edge, which is due to vignetting. There is also
slight asymmmetry in the spatial distribution of the 
diffuse background, which is caused by a small angular misalignment
of the detector plane. The count rate for diffuse background
was approximately 20~\% of the total background.

The instrument background intensity and spectrum are highly
position dependent, with a steep increase near the edges at
all radial directions. Leakage of the radiative calibration
sources causes residual line emission in the neighbourhood
of the source positions. The count rate for the instrument
background was approximately 80~\% of the total background.

The total background level varied with a range of approximately
5~\% between different orbital sections. the variation is significant, but small. Also, it is impossible to say, what fraction of this, if any, is
caused by the simultaneous variation of the solar aspect angle of the
satellite, and the unknown variations of particle radiation level.
We plan to separate these effects by the support of
future background observations.

Although our modelling is simple, and does not provide accurate
absolute estimates of physical background fluxes, it yields information
which can be applied to the JEM-X analysis software to properly
account for background contribution in spatially resolved spectral
data. A thorough analysis of JEM-X background will be presented
in a future paper.

\begin{acknowledgements}
Authors from the Observatory, University of Helsinki acknowledge
the Academy of Finland, TEKES, and the Finnish space research programme
ANTARES for financial support in this research. J. Schultz
is grateful for the financial support of the Wihuri Foundation.
The Danish Space Research Institute acknowledges support given to the
development of the JEM-X instrument from the PRODEX programme.
\end{acknowledgements}

\end{document}